%% file: main.tex
\PassOptionsToPackage{final}{graphicx}
\PassOptionsToPackage{final}{listings}
\documentclass[paper]{llncs}
\usepackage[title]{appendix}
\usepackage{sansmath}
\usepackage{amssymb}
\usepackage[ruled,vlined]{algorithm2e}
\usepackage[inline]{enumitem}
\usepackage{amsmath}
\usepackage{amssymb}
\usepackage{amsfonts}
\usepackage{amsopn}
\usepackage{amscd}
\usepackage{hyperref}
\usepackage{verbatim}
\usepackage{tikz}
\usepackage{proof}
\usepackage{alltt}
\usepackage{color}
\usepackage{xspace}
\usepackage{tikz,ifthen}
\usepackage{fancybox}
\usepackage{wrapfig}
\usepackage{multirow}
\usepackage{listings}
\usepackage{rotating}
\usepackage{xcolor,colortbl}
\usepackage[textsize=small,colorinlistoftodos]{todonotes}
\usepackage[outdir=./images/]{epstopdf}
\usepackage{sansmath}
\usepackage{mathtools}
\usepackage{uppaal}
\usepackage{stmaryrd}
\usepackage{wasysym}
\usepackage{mathtools}
\usepackage{cite}
\usepackage{proof}

\usepackage{stackengine}

\usepackage{caption}
\usepackage{subcaption}
\input{macros}

\title{It's Time to Play Safe:\\ Shield Synthesis for Timed Systems}
\author{Roderick Bloem\inst{1}  \and Peter Gj{\o}l Jensen\inst{3} \and Bettina Könighofer\inst{1,2}  \and Kim Guldstrand Larsen\inst{3} \and Florian Lorber\inst{3}
\and Alexander Palmisano\inst{1}}

\institute{
Graz University of Technology, Institute IAIK, Austria
\and
Silicon Austria Labs, TU-Graz SAL DES Lab, Austria
\and
Department of Computer Science, Aalborg University, Denmark
}

\begin{document}

\maketitle

\begin{abstract}
Erroneous behaviour in safety critical real-time systems may inflict serious consequences.
In this paper, we show how to synthesize \emph{timed shields} from timed safety properties given as timed automata.
 A timed shield enforces the safety of a
running system while interfering with the system as little as possible.
We present \emph{timed post-shields} and \emph{timed pre-shields}.
A timed pre-shield is placed \emph{before} the system and provides a set of safe outputs. This set restricts the choices of the system.
A timed post-shield is implemented \emph{after} the system. It monitors the system and corrects the system's output only if necessary.
We further extend the timed post-shield construction to provide a guarantee on the recovery phase, i.e., the time between a specification violation and the point at which full control can be handed back to the system.
In our experimental results, we use timed post-shields to ensure the safety in a reinforcement learning setting for controlling a platoon of cars, during the learning and execution phase,
and study the effect.

\end{abstract}

\section{Introduction}
\label{sec:intro}
\input{intro}

\section{Specification Theory for Real-Time Systems}
\label{sec:def}
\input{definitions}

\section{Timed Post-Shields}
\label{sec:ts}
\input{ts}


\section{Timed Pre-Shields}
\label{sec:preempt}
\input{preempt}


\section{Experiments}
\label{sec:exp}
\input{experiments}

\section{Conclusion}
\label{sec:conc}
\input{conclusion}

\bibliographystyle{abbrv}
\bibliography{references}

\begin{subappendices}
\renewcommand{\thesection}{\Alph{section}}
\section{Reinforcement Learning Configuration}\label{app:RL}
Our hyper-parameters for the DQN were chosen in the following way.
The input features consists of the distances between the cars and the velocities of the cars.
Therefore, for $n$ follower cars in the platoon, the input layer has the size $2*n+1$.
We have DNNs for actor and critic, containing 3 hidden layers with Rectified Linear Units and a linear layer for  the  output.  Networks  were  optimized  with  an  Adamax optimizer.  We  used  16  units  in  the hidden layers. We used the learning rate $\alpha= 0.002$ and
the exponential decay rates $\beta_1=0.9$ and $\beta_2=0.9999$.
The output layer is $3^{n}$, since the RL agent can pick one of three different possible accelerations
for each follower car.
The reward function is designed such that the total spacing between the vehicle
is minimized. If the distance between any two cars is either $\leq 5$m or $\geq 200$m, then the reward is set to $-1$. In all other cases, the distances between the cars are used within a logarithmic scale
to determine the reward $0\leq r\leq 1$ per step.

\end{subappendices}

\end{document}

%% file: macros.tex
\newcommand{\R}{\ensuremath{\mathbb{R_{\geq 0}}}}

\newcommand{\constraints}{\ensuremath{\mathcal{C}}}
\newcommand{\cconstraints}{\ensuremath{\mathcal{B}}}

\newcommand{\loc}{\ensuremath{\ell}}

\newcommand{\taactions}{\ensuremath{\Sigma}}

\newcommand{\vals}[1]{\mathcal{V}(#1)}

\newcommand{\iaction}[1]{\ensuremath{a?}}

\newcommand{\graphover}[1][\adtree]{\ensuremath{\mathtt{Graph}(\adtree)}}

\DeclareMathAlphabet{\mathpzc}{OT1}{pzc}{m}{it}
\def\A{\ensuremath{\mathcal{A}}\xspace}
\def\Q{\ensuremath{\mathcal{Q}}\xspace}

\newcommand\clocks[1][]{\ensuremath{{\mathit{X}_{#1}}}\xspace}

\newcommand\locs[1][]{\ensuremath{L_{#1}}\xspace}
\newcommand\invars[1][]{\ensuremath{I_{#1}}\xspace}

\newcommand\trans[1]{\ensuremath{\overset{#1}{\longrightarrow}}\xspace}

\newcommand\val[0]{\ensuremath{\nu}\xspace}

\newcommand{\edges}{\ensuremath{\mathit{E}}}

\newcommand{\implementation}{implementation\xspace}

\newcommand{\sys}{\uppProc{Sys}\xspace}
\newcommand{\spec}{\uppProc{Spec}\xspace}

\newcommand{\shield}{\uppProc{Ctr}\xspace}
\newcommand{\mspec}{\uppProc{mSpec}\xspace}
\newcommand{\specp}{\uppProc{Spec}'\xspace}
\newcommand{\mspecp}{\uppProc{mSpec}'\xspace}
\newcommand{\sh}{\uppProc{Sh}\xspace}

\newcommand{\shs}{\uppProc{Sh$_S$}\xspace}
\newcommand{\shg}{\uppProc{Sh$_G$}\xspace}
\newcommand{\shgt}{\uppProc{Sh$_{GT}$}\xspace}

\newcommand{\fspec}{\uppProc{Spec$^{f}$}\xspace}
\newcommand{\fspecone}{\uppProc{Spec$^{f_1}$}\xspace}
\newcommand{\fspecn}{\uppProc{Spec$^{f_n}$}\xspace}
\newcommand{\fspeci}{\uppProc{Spec$^{f_i}$}\xspace}

\newcommand{\mfspec}{\uppProc{mSpec$^{f}$}\xspace}
\newcommand{\mfspecone}{\uppProc{mSpec$^{f_1}$}\xspace}

\newcommand{\mfspeci}{\uppProc{mSpec$^{f_i}$}\xspace}

\newcommand{\wrm}{\uppProc{Wr}\xspace}
\newcommand{\act}{\uppProc{Act}\xspace}

\newcommand{\swr}{\mathcal{S}_{wr}\xspace}

%% file: intro.tex
Today's systems are becoming increasingly sophisticated and powerful. 
At the same time, systems have to perform highly safety critical tasks, e.g., in the
domain of self-driving cars, and make extensive use of machine learning, such as
reinforcement learning~\cite{FultonP19}.
As a result, complete offline verification 
is rarely possible. 
This holds especially true for safety critical real-time systems, where a deadline violation comes with serious consequences.
An alternative is runtime verification~\cite{BlechFB12, BauerLS11, BartocciFBCDHJK19, BartocciFFR18}. 
Runtime enforcement  (RE)~\cite{KonighoferABHKT17, WuZW16, FalconeMFR11} extends this by enforcing the expected behavior of a system at runtime.

\begin{figure}
\captionsetup[subfigure]{font=footnotesize}
\centering
\subcaptionbox{Timed post-shielding}[.5\textwidth]{%
\begin{tikzpicture}
\definecolor{myblack}{cmyk}{.67,.33,0,.99}
\tikzstyle{box} = [rectangle, rounded corners, minimum width=1.5cm, minimum height=0.7cm, text centered, draw=myblack]
\tikzstyle{arrow} = [thick,->,>=stealth]
\tikzstyle{point} = [coordinate]
\definecolor{dgreen}{cmyk}{60, 0,100,0}

\node (system) [box, xshift=1.5cm] {$Sys$};
\node[inner sep=1pt, right of=system, xshift=2cm] (shield) {{{\includegraphics[bb= 15 10 800 780,scale=0.07]{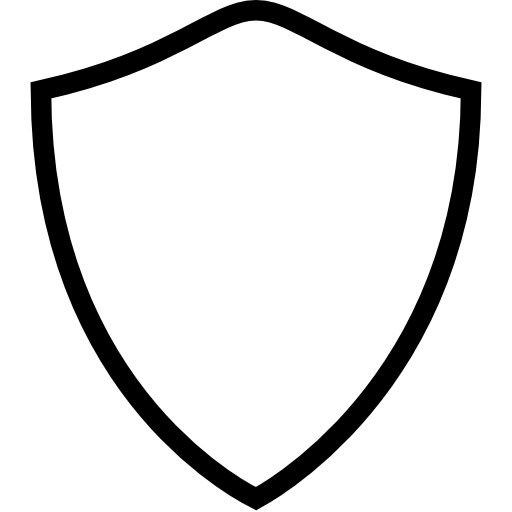}}}};

\draw[arrow, thick] (0,0) --  (system.west);
\draw[thick] (0.3,0) --  (0.3,-0.7);
\draw[arrow, thick] (0.3,-0.7) -- node [pos=0.75, above]{input} (3.8,-0.7);

\draw [arrow, thick] (system) -- node [anchor=south]{output} (shield);

\draw [arrow, thick] (4.6,-0.4) -- (6,-0.4);

\node[text width=1cm] at (4.45,-0.3) {$Sh$};
\node[text width=1cm] at (5.3,-0.4) {safe output};
\end{tikzpicture}}%
\subcaptionbox{Timed pre-shielding}
[.5\textwidth]{\begin{tikzpicture}
\definecolor{myblack}{cmyk}{.67,.33,0,.99}
\tikzstyle{box} = [rectangle, rounded corners, minimum width=1.5cm, minimum height=0.7cm, text centered, draw=myblack]
\tikzstyle{arrow} = [thick,->,>=stealth]
\tikzstyle{point} = [coordinate]
\definecolor{dgreen}{cmyk}{60, 0,100,0}

\node[inner sep=1pt, xshift=1.7 cm, yshift=0] (shield) {{{\includegraphics[bb= 15 10 800 780,scale=0.07]{images/shield-icon-blank.png}}}};
\node (system) [box, right of=shield, xshift=1.3cm, yshift=-0.3cm] {$Sys$};

\draw[arrow, thick] (0,-0.3) --  (0.8,-0.3);
\draw[thick] (0.3,-0.3) --  (0.3,-1.1);
\draw[thick] (0.3,-1.1) -- node [pos=0.75, above]{input} (2.8,-1.1);
\draw [thick] (2.8,-1.1) -- (2.8,-0.5);
\draw [arrow, thick] (2.8,-0.5) -- (3.25,-0.5);

\draw [arrow, thick] (1.8,-0.3) -- (system);

\draw [arrow, thick] (4.75,-0.3) -- node [pos=0.6, above]{output} (5.7,-0.3);

\node[text width=1cm] at (1.65,-0.3) {$Sh$};

\node[text width=1.1cm] at (2.6,0.2) {safe outputs};

\end{tikzpicture}}
\caption{Types of shielding.}
\label{fig:attach_shield}
\vspace{-5pt}
\end{figure}

In this paper, we focus on the enforcement of regular timed properties for reactive systems
and automatically synthesize \emph{timed shields} from \emph{timed automata} specifications.
A timed shield can be attached to a system in two alternative ways.
A \emph{timed post-shield} (see Fig.~\ref{fig:attach_shield}(a)) is implemented after the system.
It monitors the system and corrects the
system's output if necessary. A \emph{timed pre-shield} (see Fig.~\ref{fig:attach_shield}(b)) is placed before the system and provides a list of safe outputs to the system. This list restricts the choices of the system to safe actions. 

Timed post-shields guarantee the following two properties.
(1) \emph{Correctness:} the shielded system satisfies the safety specification, and
(2) \emph{No-Unnecessary-Deviation:}
the shield intervenes with the system only if safe system behavior would be
endangered otherwise. 
We extend timed post-shields to shields that additionally provide guarantees on the recovery time, after which control is handed back to the system; i.e., from that point on the shield forwards the outputs to the environment and does not deviate anymore.
(3a) \emph{Guaranteed-Recovery:}
the recovery phase ends within a \emph{finite time}, and
(3b) \emph{Guaranteed-Time-Bounded-Recovery:}
the recovery phase ends after a given \emph{bounded time}.

Timed pre-shields guarantee the following two properties.
(1) \emph{Correctness:} the shield provides only safe outputs to the system, and
(2) \emph{No-Unnecessary-Restriction:}
the shield provides all safe outputs to the system.

\begin{figure}[tb]
\centering
\begin{tikzpicture}
\definecolor{myblack}{cmyk}{.67,.33,0,.99}
\tikzstyle{box} = [rectangle, rounded corners, minimum width=2.5cm, minimum height=0.9cm, text centered, draw=myblack]
\tikzstyle{arrow} = [thick,->,>=stealth]
\tikzstyle{point} = [coordinate]
\definecolor{dgreen}{cmyk}{60, 0,100,0}

\node (env) [box] {Environment};
\node (la) [box, right of=env, xshift=3.4cm] {Learning Agent};


\node[inner sep=1pt, right of=env, xshift=3.8cm, yshift=-1.2cm] (shield) {{{\includegraphics[bb= 15 10 800 780,scale=0.07]{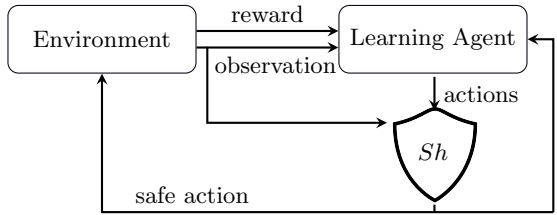}}}};

\draw [arrow, thick] (env.5) -- node [anchor=south]{reward} (la.175);
\draw [arrow, thick] (env.-5) -- node [anchor=north]{~~observation} (la.185);

\draw [arrow, thick] (1.4, -0.1) |- node [anchor=south]{} (shield.175);

\draw [arrow, thick] (6,-2.3) -| node[pos=0.4, above] {safe action} (env);
\draw [arrow, thick] (6,-2.3) |- node[anchor=north] {} (la);

\draw [arrow, thick] (4.42, -0.5) -- node [anchor=west]{actions} (4.42, -0.95);
\node[text width=1cm] at (4.7,-1.5) {$Sh$};
\draw[thick] (4.42, -2.2) --  (4.42,-2.3);

\end{tikzpicture}
    \caption{A timed post-shield in a reinforcement learning setting.}
    \label{fig:attach_postposed_shield_learner}
\end{figure}

Shields can be employed during reinforcement learning (RL)
to ensure safety by enforcement both during a system's learning and execution phases~\cite{10.1007/978-3-319-11936-6_10,AlshiekhBEKNT18}. 
We introduce a timed post-shield after the learning agent, as depicted in
Fig.~\ref{fig:attach_postposed_shield_learner}. 
The shield monitors the actions selected by the learning agent and corrects them if and only if the chosen action is unsafe.

To sum up, we make the following contributions:
\begin{itemize}
  \item We propose to synthesize timed shields from timed automata specifications.
  \item We discuss two basic types of timed shields: pre-shields and post-shields.
  \item We propose timed post-shields with the ability to recover.
  \item Our experiments show the potential of timed shields to enforce safety in RL.
\end{itemize}

\textbf{Related Work.}
In most work about RE, an enforcer monitors a program
that outputs events and can either terminate the program once it detects an error~\cite{Schneider00},
or alter the event in order to guarantee, for example, safety~\cite{HamlenMS06} or privacy~\cite{JiL17, WuRRLS18}.
Renard et al.~\cite{RenardFRJM19} and Falcone et al.~\cite{FalconeP19} considered RE for timed properties. The similarities between these enforcers and a shield is in
their ability to alter events. Their work only considers static programs whereas we consider enforcing correctness for reactive systems.
 
The term runtime assurance~\cite{Sha01} is often used if there exist 
a switching mechanism that alternates between running a high-performance system
and a provably safe one. These concept is similar to post-shielding, with the 
difference that we synthesize the entire provable safe system (the shield) including the switching mechanism while proving guarantees on recovery.

Könighofer et. al. introduced ~\cite{KonighoferABHKT17} shield synthesis from LTL specifications, which, while related, are not expressible enough to capture real time behavior and thus cannot shield against timing related faults in real timed systems.
Wu et al.~\cite{WuWDW19} extend shields for boolean signals to real-valued shields to enforce the safety of cyber-physical systems.
The concept of shield synthesis for RL from (probabilistic) LTL specification was discussed in~\cite{AlshiekhBEKNT18, abs-1807-06096} and early work on combining (basic) shielding with RL was demonstrated in~\cite{10.1007/978-3-319-11936-6_10}. 

We use the specification theory of Timed  In-put/Output Automaton (TIOA) used by \uppaalecdar~\cite{david2010ecdar}. 
To synthesize our timed shields we use \uppaaltiga~\cite{CassezDFLL05}, a tool which implements algorithms for solving games based on timed game automata with respect to reachability and safety properties, producing non-deterministic safety strategies. \uppaalstratego~\cite{DavidJLMT15, 10.1007/978-3-319-11936-6_10}
extends \uppaaltiga by the capability of optimizing these strategies with respect to desired performance measures.
While \uppaalstratego contains a RL component, we utilize a third-party, off-the-shelf, RL-system to demonstrate the applicability of the proposed method in a generic RL setting.

\textbf{Outline.} In Sec.~\ref{sec:def}
we give the formal notation and constructions.
Definition and construction of
timed post-shields (with recoverability garuantees under different fault models) are introduced in Sec.~\ref{sec:ts}  (Sec.~\ref{sec:tsg}). 
Timed pre-shields are introduced in Sec.~\ref{sec:preempt}.
We discuss our experimental findings from a car platooning problem in Sec.~\ref{sec:exp} followed by a conclusion in Sec.~\ref{sec:conc}.

%% file: definitions.tex

Let us recall definitions of Timed (Input/Output) Automata.
%
Let $\clocks$ be a finite set of real-valued \emph{clocks}.
Let $\vals{\clocks}\mapsto\R$ be the valuations over $\clocks$ and let $\overset{\rightarrow}{0}$ be the valuation that assigns $0$ to each clock.
For the value of a single clock of a given valuation $\val\in\vals{\clocks}$, we write $\val(\uppClock{x})$ and $\val[\uppClock{x}]$ denotes the reset of the clock $\uppClock{x}$ in the valuation $\val$.
We extend the notion of reset to sets, i.e., for some $Y\subseteq \clocks$ let $\val[Y]$ be the valuation after resetting all values of clocks in $Y$ to zero
and otherwise retaining their value.
If $\delta\in \R$ is a positive delay, then we denote by $\val+\delta$ the valuation s.t. for all $\uppClock{x}\in X$, $(\val + \delta)(\uppClock{x})=\val(\uppClock{x})+\delta$.

Let $Y \subseteq \clocks $ and $Z \subseteq \clocks \cup \mathbb{Z} $. We denote the set of all simple constraints as $\Phi(Y, Z)=Y\times\{\uppProp{<,\leq,\geq, >}\}\times Z$.
The set of all  possible clock constraints is defined by $\constraints(\clocks)=2^{\Phi(\clocks, \clocks\cup\mathbb{Z})}$.
We denote the (conjunctive) subset of restricted clock constraints by $\cconstraints(X)\subseteq\constraints(X)$ with $\cconstraints(X)= 2^{\Phi(\clocks,\mathbb{Z})}$.

\begin{definition}[Timed Input/Output Automaton (TIOA)~\cite{alur1994theory, david2010timed}]
A Timed Input/Output Automaton (TIOA) is a tuple $\A=(\locs,\loc_0,\taactions{}^?,\taactions{}^!,\clocks,\edges,\invars)$ where
	$\locs$ is a finite set of locations,
	$\loc_0$ is the initial location,
	$\taactions{}=\taactions{}^?\cup\taactions{}^!$ is a finite
   set of actions partitioned into inputs ($\taactions{}^?$) and outputs ($\taactions{}^?$),
	 $\clocks$ is a set of clocks,
	 $\edges\subseteq \locs\times\cconstraints(\clocks)\times(\taactions{}^!\cup\taactions{}^?)\times2^{\clocks} \times \locs$ 
	 is a set of edges, and 
	$\invars:\locs\rightarrow\cconstraints(\clocks)$ is a function assigning invariants to locations.
\end{definition}
States of a TIOA are given as a pair $(\loc,\val)\in \locs\times\mathbb{R}_{\geq_0}^{\clocks}$ consisting of a discrete location and a real-valued assignment to the clocks.
From a given state $(\loc,\val)\in \locs\times\mathbb{R}_{\geq_0}^{\clocks{}}$ where $\val\models\invars(\loc)$, we have two kinds of transitions; (1) \emph{discrete transitions}: $(\loc,\val)\trans{\alpha}(\loc',\val')$ if there exists $(\loc,\psi,\alpha,Y,\loc')\in \edges$ s.t. $\val\models \psi$, $\val'=\val[Y]$ and $\val'\models\invars(\loc')$, and
	(2) \emph{delay transitions} for some $\delta\in \R$ where $(\loc,\val)\trans{\delta}(\loc,\val')$ if $\val'=\val+\delta$ and $\val'\models\invars(\loc)$.
We define the semantics of a TIOA as a TLTS.
\begin{definition}[Timed Labeled Transitions System (TLTS)~\cite{alur1994theory, david2010timed}]
A TLTS is a tuple $\llbracket \A \rrbracket=(\Q,q_0,\rightarrow)$ s.t. $\Q=\locs\times\mathbb{R}_{\geq_0}^{\clocks{}}$, $q_0=(\loc_0,\overset{\rightarrow}{0})$,
and $\rightarrow: Q \times \taactions^!{}\cup\taactions^?{}\cup\R \times Q $ is a transition relation defined  as above.
\end{definition}
A {\em trace} $\sigma$ of an TIOA is a finite sequence of alternating delay and discrete transitions of the form
$(l_{0}, v_{0}) \xrightarrow{\delta_1} (l_{0}, v_{0} + \delta_{1}) \xrightarrow{\tau_{1}} (l_{1}, v_{1}) \xrightarrow{\delta_{2}} \cdots
\xrightarrow{\delta_{n}} (l_{n-1}, v_{n-1} + \delta_{n}) \xrightarrow{\tau_{n}} (l_{n}, v_{n})$,
where $v_{0} = \overset{\rightarrow}{0}$ and $\tau_{i} = (l_{i-1}, \alpha_{i}, g_{i},  Y, l_{i}) \in \edges$. By agreement we let $\trans{\ast}$ denote the transitive closure of the transition-function.


The state space of TIOA can be represented via \emph{zones}~\cite{alur1994theory}, symbolic sets of states containing the max. set of clock valuations satisfying given constraints. 

In this paper, we only consider deterministic and input enabled TIOA.
$\A$ is \textit{deterministic}, if $\llbracket \A \rrbracket$ satisfies that $\forall \alpha \in \taactions{}, \forall (\loc,\val) \in \Q . (\loc,\val)\trans{\alpha}(\loc',\val') \wedge (\loc,\val)\trans{\alpha}(\loc'',\val'') \implies \loc' = \loc'' \wedge \val' = \val''$. Thus, in every state,  transitions of a given label will lead to a unique state. 
$\A$ is \textit{input-enabled}, if $\llbracket \A \rrbracket$ satisfies that $\forall \alpha \in \taactions{}^?, \forall (\loc,\val) \in \Q . (\loc,\val)\trans{\alpha}(\loc',\val')$. 
Thus, every state can receive any input.  For simplicity, we assume implicit input-enabledness in our figures.

TIOA can be used as \emph{specifications}. 
Specifications are not necessarily exact on timing behaviour (e.g., on the output of a label) but allow for ranges of timing. 
An \emph{implementation} satisfies a specification as long as any behavior is included in that of the specification.

\begin{figure}[tb]
	
	\minipage{0.49\textwidth}
	\centering
	\includegraphics[width=0.7\linewidth]{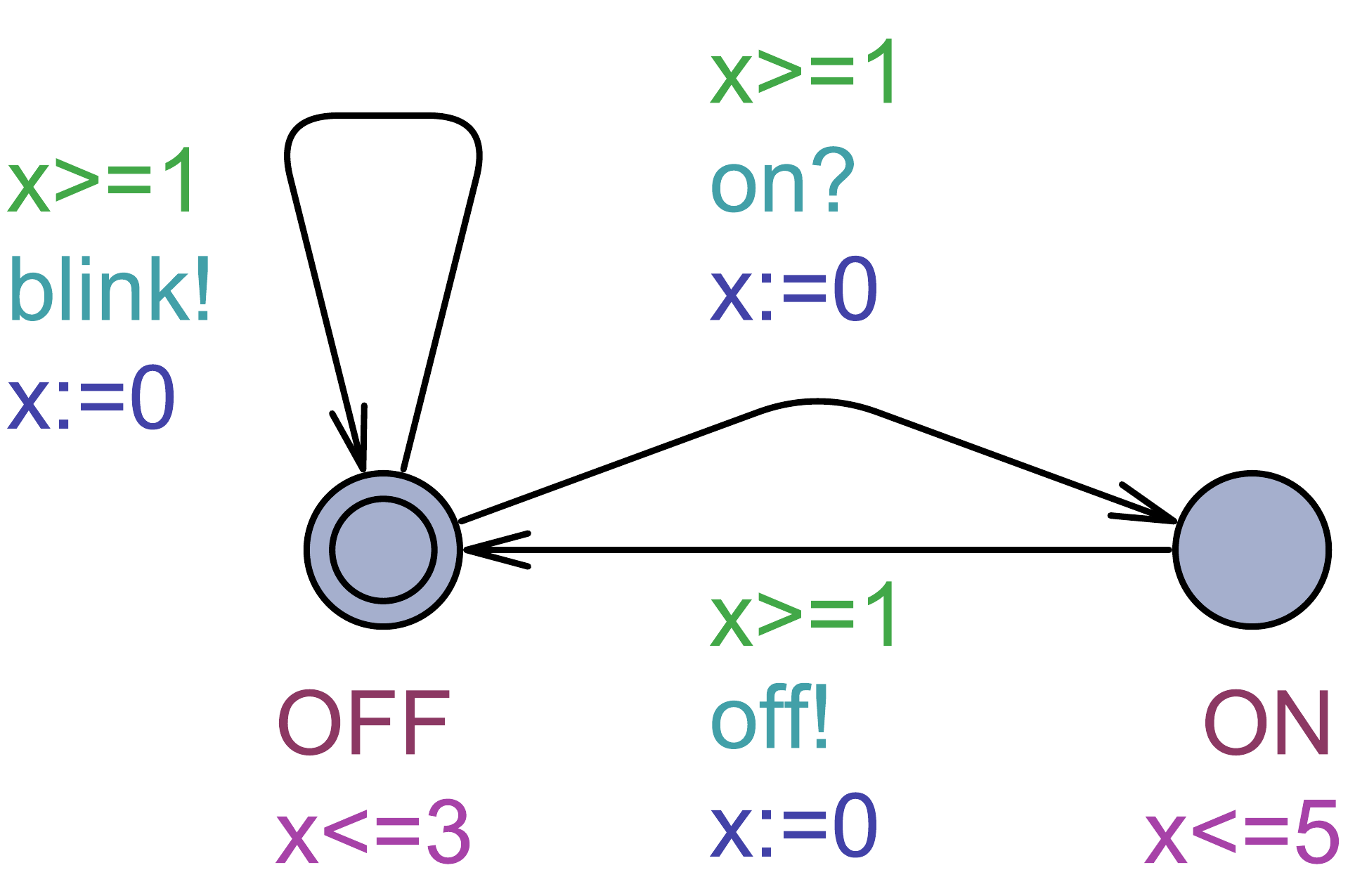}\\
	(a)
	\endminipage\hfill
	\minipage{0.49\textwidth}
	\centering
	\includegraphics[width=0.7\linewidth]{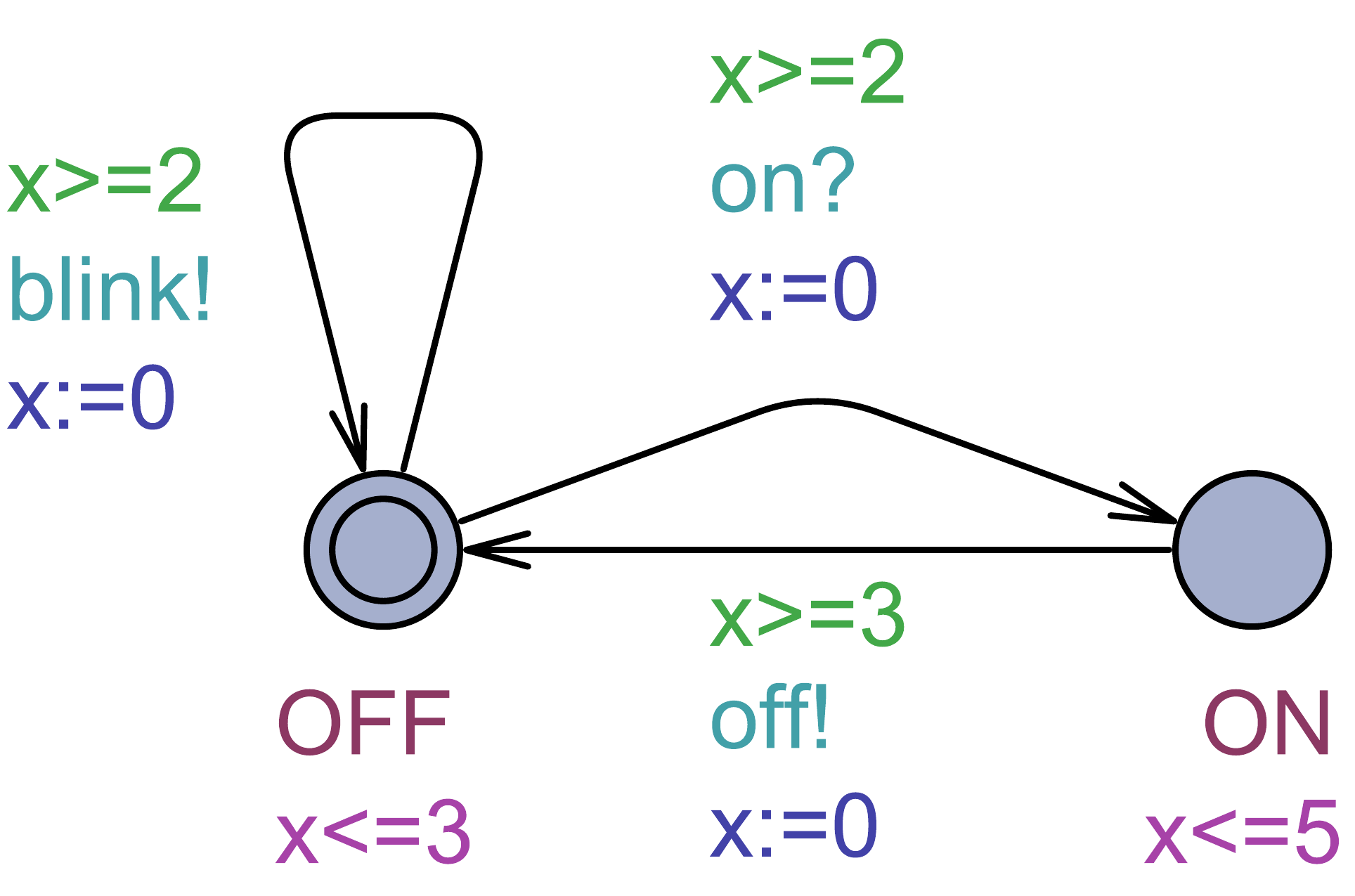}\\
	(b)
	\endminipage\hfill
	\caption{Specification $\spec_1$ (a) of a light switch; specification $\spec_2$ refines it (b).}
	\label{fig:example_kim}
	\vspace{-10pt}
\end{figure}

\textbf{Example.} \emph{Fig.~\ref{fig:example_kim}(a) shows a specification  $\spec_1$ of a light switch: whenever the light is switched \uppLoc{ON},  this setting has to be kept for $1$ to $5$ time units. Whenever the light is switched \uppLoc{OFF}, the light switch has to blink at least once every $3$ time units. The timing is tracked via the clock $\uppClock{x}$.}

We say that a TIOA $\A_\mathcal{I}$ \emph{refines} a TIOA $\A_\mathcal{S}$ if the corresponding TLTS $\llbracket \A_\mathcal{I} \rrbracket$ refines the TLTS $\llbracket\A_\mathcal{S} \rrbracket$, let us formally defined this relationship.

\begin{definition}[Refinement\cite{david2010timed}] \label{def:ref}
A TLTS $\mathcal{I}=(\Q_\mathcal{I},q_0^\mathcal{I},\rightarrow_\mathcal{I}) $ refines a TLTS $\mathcal{S}=(\Q_\mathcal{S},q_0^\mathcal{S},\rightarrow_\mathcal{S})$, written $\mathcal{I} \leq \mathcal{S}$, iff there exists a binary relation $R\subseteq \Q_\mathcal{I} \times \Q_\mathcal{S}$ containing $(q_0^\mathcal{I},q_0^\mathcal{S})$ such that for each pair of states $(q_\mathcal{I},q_\mathcal{S}) \in R$ we have:
\begin{enumerate}
\item if $\exists q'_\mathcal{S} \in \Q_\mathcal{S}. ~q_\mathcal{S}  \xrightarrow{i?}_\mathcal{S} q'_\mathcal{S}$
 then $\exists q'_\mathcal{I} \in \Q_\mathcal{I}. ~
  q_\mathcal{I}  \xrightarrow{i?}_\mathcal{I} q'_\mathcal{I}$ and $(q'_\mathcal{I},q'_\mathcal{S}) \in R$
\item if $\exists q'_\mathcal{I} \in \Q_\mathcal{I}. q_\mathcal{I}  \xrightarrow{o!}_\mathcal{I} q'_\mathcal{I}$ then $\exists q'_\mathcal{S} \in \Q_\mathcal{S}. q_\mathcal{S}  \xrightarrow{o!}_\mathcal{S} q'_\mathcal{S} $ and $(q'_\mathcal{I},q'_\mathcal{S}) \in R$ and 	
\item if $\exists \delta \in \mathbb{R}_{\geq 0}. q_\mathcal{I}  \xrightarrow{\delta}_\mathcal{I} q'_\mathcal{I}$ then $q_\mathcal{S}  \xrightarrow{\delta}_\mathcal{S} q'_\mathcal{S}$ and $(q'_\mathcal{I},q'_\mathcal{S}) \in R$.

\end{enumerate}
\end{definition}

\textbf{Example.} \emph{Fig.~\ref{fig:example_kim}(b) shows a specification  $\spec_2$ 
that refines the specification $\spec_1$ of Fig~\ref{fig:example_kim}(a) by restricting the timings of some of the signals.}

Given a specification \spec and a system \sys, a \textbf{monitor} \mspec observes
\sys w.r.t \spec (i.e., all transitions in \mspec are inputs) and enters an
error-state \uppLoc{ERR} whenever non-conformance between \sys and \spec is observed, that is, whenever
an output of \sys is observed that is not allowed by \spec.
\begin{figure}[tb]	
	\centering
	\includegraphics[width=0.59\linewidth]{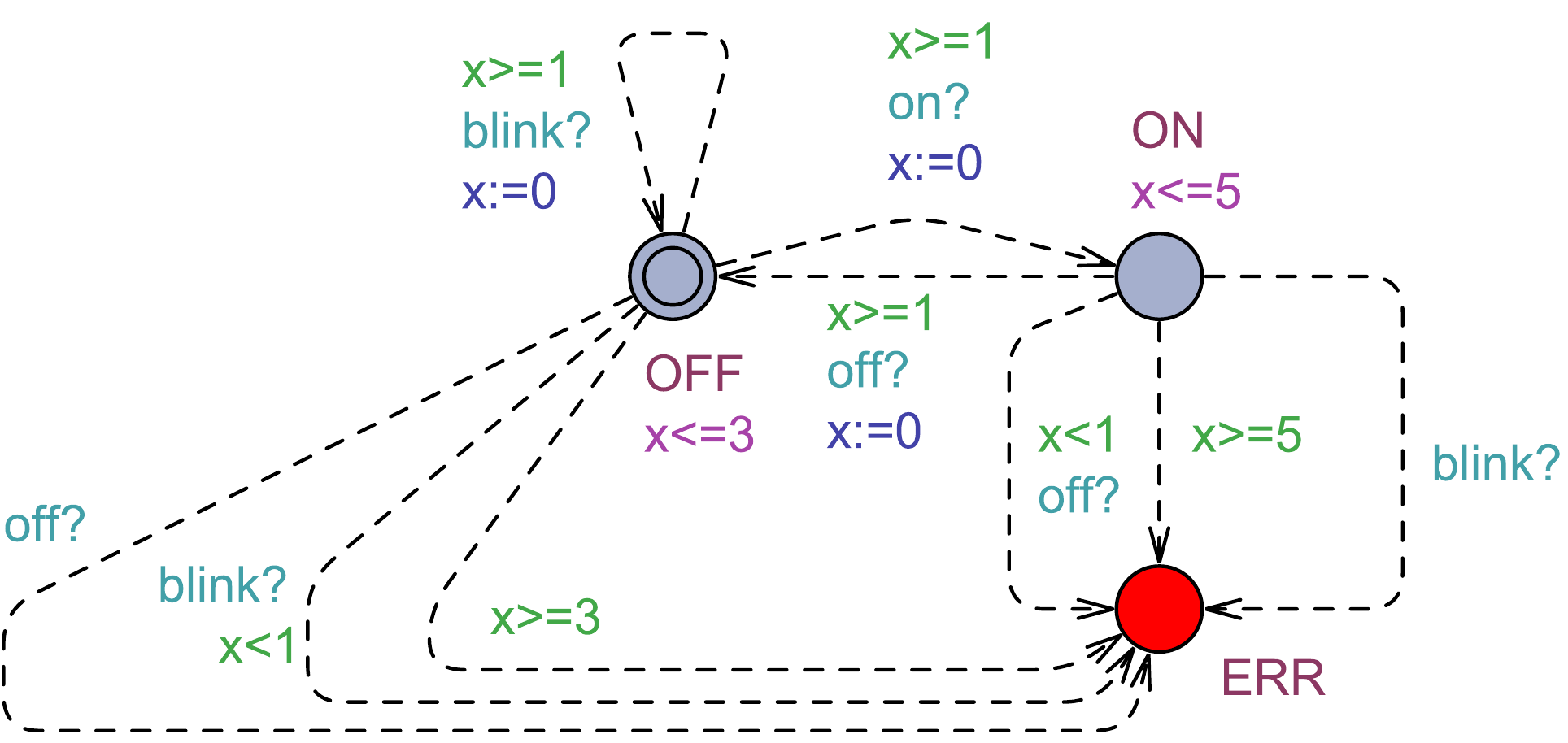}	
	\caption{A monitor \mspec of the light switch specification $\spec_1$ of Fig.~\ref{fig:example_kim}(a).}
	\label{fig:mon}
	\vspace{-10pt}
\end{figure}

\textbf{Example. }\emph{Fig.~\ref{fig:mon} depicts a monitor for $\spec_1$ of Fig.~\ref{fig:example_kim}(a).
}

We use \emph{Networks of Timed Automata} to enable parallel composition of TIOA. 
\begin{definition}[Networks of Timed Automata (NTA)~\cite{alur1994theory}]
Let $\Sigma$ be a set of actions and let $\Sigma_1,\dots,\Sigma_n$ be a partitioning of $\Sigma$. 
Let $\A_1,\A_2,\dots,\A_n$ be TIOAs, where $\llbracket \A_i \rrbracket = (\Q^i,q_0^i,\rightarrow^i)$
and $\A_i$ has $\Sigma_i$ as its output and $\Sigma$ as its input alphabet.
A network of TIOA $\A_1\|\A_2\| \dots \| \A_n$ is defined via the 
TLTS $(\Q^1\times\dots\times\Q^n,(q_0^1,... , q_0^n),\rightarrow)$ with:
$$\frac{[s_i \xrightarrow{\delta}_i s'_i]_{i=1 \dots n}}{(s_1, \dots, s_n) \xrightarrow{\delta}  (s'_1, \dots, s'_n)}~~\text{with } \delta \in \R \text{, and}$$
$$\frac{[s_j \xrightarrow{\alpha!}_j s'_j] [s_i \xrightarrow{\alpha?}_i s'_i]_{i \neq j}}{(s_1, \dots, s_n) \xrightarrow{\alpha!}  (s'_1, \dots, s'_n)}~~\text{with }{\alpha \in \Sigma_j}.$$ 
\end{definition}
For deterministic specifications, the following important theorem, given in the syntax of \uppaal queries~\cite{behrmann2004tutorial}, holds, stating that a system refines a specification iff the parallel product of the system and the specification monitor cannot reach the error-state.
\begin{theorem}
	$\forall~ \sys.~ \sys \leq \spec~~$ iff $~~(\sys ~|~ \mspec) \models A \Box~ \neg
	\mspec.\uppLoc{ERR} $
\end{theorem}

\begin{definition}[Timed Game Automaton (TGA)~\cite{maler1995synthesis}]
 A Timed Game Automaton (TGA) is a TIOA in which the set of output actions $\taactions^{!}$ is partitioned into controllable actions ($\taactions_C$) and uncontrollable actions ($\taactions_U$).
\end{definition}
The definition of NTA trivially extends to Networks of TGA (NTGA).

%
Given a TGA $\mathcal{G}=(\langle\locs,\loc_0,\taactions{},\clocks,\edges,\invars,\rangle,$ $\taactions_U,\taactions_C)$, a \emph{memoryless strategy} $\omega:\locs{}\times\mathbb{R}_{\geq_0}^{\clocks{}}\rightharpoonup 2^{\{\lambda\}\cup\taactions{}_C}$ is a  function over the states of $\mathcal{G}$ to the set of controllable actions or a special nothing-symbol $\tau$.

\begin{definition}[Strategy Composition]
Given a TGA $\mathcal{G}=(\langle\locs,\loc_0,\taactions{},\clocks,\edges,\invars,\rangle,$ $\taactions_U,\taactions_C)$ and a memoryless strategy $\omega$, the composition $\mathcal{G}\|\omega$ provides a restriction of the transition-system of $\mathcal{G}$;
\begin{itemize}
	\item if $(\loc,\val)\trans{\alpha}(\loc',\val')$ then either we have $\alpha\in\Sigma^?$ or we have $\alpha\in\Sigma^!$ and $\alpha\not\in\taactions_C$ or $\alpha\in\omega(\loc,\val)$, and
	
		\item if $(\loc,\val)\trans{\delta}(\loc',\val')$  for $\delta\in\R$ then $\forall \delta'<\delta$ it holds that $\lambda\in\omega(\loc,\val+\delta')$.

\end{itemize}
\end{definition}

Let $\varphi\subseteq\locs$ be a set of losing locations for a TGA $\mathcal{G}$. The \textit{safety control problem} consists in finding a strategy $\omega$, s.t. $\mathcal{G}\|\omega$ constantly avoids $\varphi$. A trace $\sigma= (l_{0}, v_{0}) \xrightarrow{\delta_1} (l_{0}, v_{0} + \delta_{1}) \cdots
\xrightarrow{\delta_{n}} (l_{n-1}, v_{n-1} + \delta_{n}) \xrightarrow{\tau_{n}} (l_{n}, v_{n})$ is \textit{winning} if  $\forall k \leq n : l_k \notin \varphi$. A strategy $\omega$  is winning, if all traces of $\mathcal{G}\|\omega$ are winning.

We denote by $\wrm\subseteq\locs\times\mathbb{R}_{\geq_0}$ the \emph{winning region}, i.e., the set of all states $s$ such that there exists a winning strategy from $s$. We denote by \emph{correct} (\emph{wrong}) outputs for a state $s$ the outputs that lead to an $s' \in \wrm$  ($s' \not\in \wrm$, respectively). We denote by $\swr$ the set of states  in $\wrm$, such that any delay would leave $\wrm$.

%% file: ts.tex
This section defines and gives the construction of timed post-shields, illustrated in Fig.~\ref{fig:attach_shield}(a).
A timed post-shield is attached after the system, monitors its inputs and outputs, corrects the system's output if necessary and forwards the correct output to the environment. 

\subsection{Definition of Timed Post-Shields}

In this section, we define a timed post-shield based on its two desired
properties: 


\begin{definition}[Correctness for Post-Shields.] \label{def:corr}
Let \spec be a specification, and let \sys be a timed system. We say that a shield
\sh ensures correctness if and only if it holds that 
$$(\sys~|~\sh)~ \leq \spec.$$
\end{definition}
That is, for any (faulty) system \sys, if it is placed in parallel with the shield, the shielded system is guaranteed to satisfy the specification.

\begin{definition}[No-Unnecessary-Deviation.] \label{def:no_dev}
Let \spec be a specification, let \mspec be its corresponding monitor, let \sys be a timed system, and let \sh be a shield. 
%
%
Let $\sigma$ be any \emph{correct} timed trace of \sys, i.e., every action in $\sigma$ is correct, and no state along $\sigma$ is in $\swr$.  
We say that \sh does not deviate from \sys unnecessarily, if $(\sys|\sh)$ keeps the output for $\sigma$ intact.


\end{definition}
In other words if \sys does not violate \spec, \sh simply forwards the outputs of \sys 
to the environment without altering them. Once we reach a state in $\swr$, there is no way to know whether the system would produce an output on time, and the shield is allowed to deviate.

\begin{definition}[Timed Post-Shields.]
Given a specification \spec, \sh is a timed post-shield if for any
timed system \sys, it holds that \sh enforces correctness of the shielded system 
w.r.t. \spec (Def.~\ref{def:corr}) and \sh does not deviate from \sys unnecessarily 
(Def.~\ref{def:no_dev}).
\end{definition}

\subsection{Construction of Timed Post-Shields}

In this section we discuss the synthesis procedure of timed post-shields without guarantees on the recovery time.

\textbf{Algorithm 1.} Let \spec be a specification, \mspec a monitor for \spec, and \sys a timed system.
We construct a timed post-shield \sh
via the following steps. 

\textbf{Step 1: Construction of the monitor \mspecp.}
To differentiate the outputs given by the system and those given by the shield, we prime the outputs of the shield. The monitor \mspecp is a copy of \mspec, where all outputs are primed. \mspecp is used to ensure that the outputs of the shield are correct.

\textbf{Step 2: Construction of the automaton \shield.}
The automaton \shield  is the only component that contains controllable transitions and depicts the control options for the shield, i.e., \shield produces the primed outputs. \shield is constructed such that the no-unnecessary-deviation property
is satisfied by the shield. 
Therefore, \mspec informs \shield whether the system's output is wrong or the current state is in $\swr$, i.e., the winning region of \mspec is left.
\shield can perform three types of actions: (1) \textit{pre-fault actions}: before an error was detected, \shield can only mirror actions produced by \sys; (2) \textit{post-fault actions}: after an error was detected, \shield has full control and can chose any action at a any time; and (3) \textit{last-chance actions}: whenever the current state is in $\swr$ and any delay would leave the winning region, the shield can prevent that possibility and can choose any action that is allowed in the current state of \mspecp.

\begin{figure}[tb]
	\begin{center}
		\includegraphics[width=0.5\linewidth]{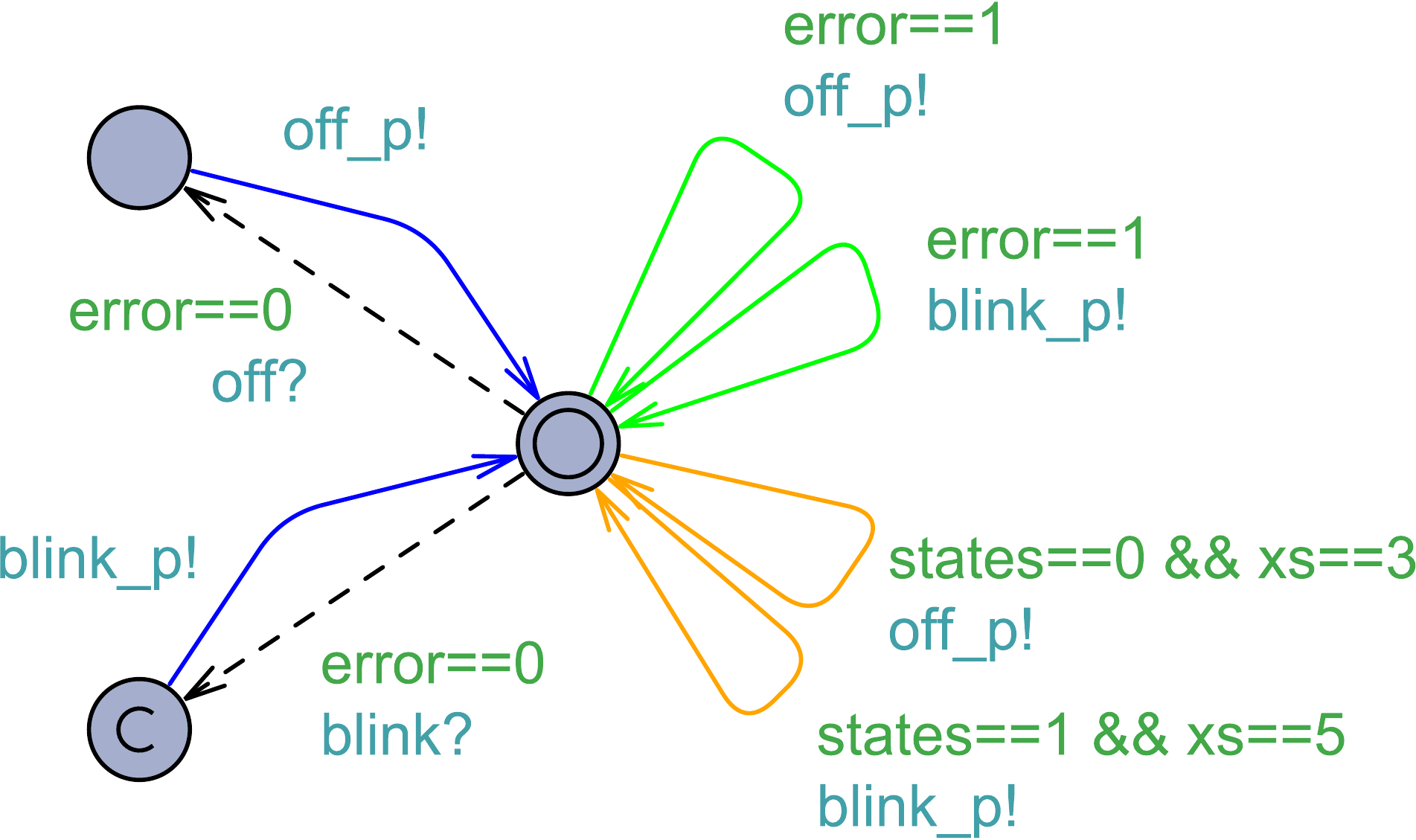}
		\caption{An automaton \shield,  defining control options of a shield for the light switch.}
		\label{fig:ex_shield}
	\end{center}
	\vspace{-10pt}
\end{figure}

\textbf{Example.} \emph{Fig.~\ref{fig:ex_shield} depicts the \shield component for the light switch of Fig.~\ref{fig:example_kim}(a). 
The two edges on the left (marked in blue) are pre-fault actions, that copy the system behaviour if no fault was detected, i.e., \uppGuard{error==0} (The error variable is set when \sys produces an output that would leave the winning region.).
The two transitions in the top right corner (green) are post-fault actions: if an error was detected, i.e., \uppGuard{error==1},  they will be enabled, and \shield can choose any output. 
The two transitions in the bottom
right (orange) show last-chance actions, for the case where the specification reached a time bound, in which case
\shield is allowed to take over, as a fault of the system may be imminent.}

\textbf{Step 3: Construct the timed safety game $\mathcal{G}$.}
We construct the timed game $\mathcal{G}$ by the following composition:
$$\mathcal{G} = ~\mspec~|~\mspecp~ |~\shield$$

In this game the monitor
\mspec observes whether the system \sys satisfies the specification \spec,
the monitor \mspecp observes whether the outputs of the shields are correct,
and \shield enforces that the shield does not deviate unnecessarily and is in charge of producing the primed outputs. 

\textbf{Step 4: Compute a strategy $\omega_S$ of $\mathcal{G}$.}
Post-shields ensure correctness (safety), i.e., 
the control objective is to ensure that \specp is
never violated by the shielded system. This can be expressed via the following safety 
query specifying that the error state should never be reached, given in \uppaaltiga syntax.
$$
\uppProp{control: A \Box~ \neg \mspecp.\uppLoc{ERR}}
$$

Solving the safety game w.r.t. this query produces a strategy $\omega_S$, which 
we use in the next step to produce a timed post-shield. 

\textbf{Step 5: Construction of the timed post-shield \shs.}
From the strategy $\omega_S$, we construct the shield \shs in the following way:
A shield \shs is the network of timed automata received by composing $\mathcal{G}$ with the derived strategy $\omega_S$, denoted by $\mathcal{G}\|\omega_S$, meaning that all unsafe transitions (or transitions that would not lead to recovery) are restricted. This shield may still permit multiple outputs in a given state, any of which ensures safety.

\begin{theorem}
A shield \shs constructed according to Alg. 1 is a timed post-shield.
\end{theorem}
\emph{Proof}. We have to proof that \shs satisfies the correctness property and the 
no-unnecessary-deviation property. 
\shs satisfies the correctness property, since all transitions 
leaving the winning region were removed from $\mathcal{G}|\omega_S$. It thus holds that $(\sys~|~\sh)~ \leq \specp$.
Thus the primed outputs of the shield satisfy the specification.
Additionally, the construction $\mathcal{G}$ via the automaton \shield ensures that 
the shield cannot alter an action before a fault occurs, thereby ensuring that 
the shield cannot deviate unnecessarily \qed


\section{Timed Post-Shields with Recovery Guarantees}
\label{sec:tsg}

In this section, we first discuss the challenges that we face when
synthesizing shields with the ability to recover from system faults
and discuss assumptions that we make on the system and its faults that are necessary for
our synthesis procedure.
Next, we define and construct timed post-shields with 
\textit{guaranteed recovery} and \textit{guaranteed time-bounded recovery}.

\subsection{Recovery under Fault Models}

In shield synthesis, we consider the system to be shielded as a black box,
which brings a huge scalability advantage, especially when shielding complex timed implementations.
Therefore, a shield has to ensure correctness for any system.

In order to end the recovery phase and to hand back control to the system, the shield
needs to \textit{resolve} the error that occurred, i.e., the state of the shield needs to align with the actual state of the system. The tricky part is, as mentioned before, that the system is considered as a black box and the shield can only observe the system's outputs but not its internal state.

In this paper we assume that the only violations that happen are due to {\em transient errors}. 
\begin{definition}[Transient Error]
A transient error is an error that happens only once, and has correct pre-error and post-error behavior.
\end{definition}
To determine the state of a system in case of a fault, we launch several \emph{fault models}
and assume that one of this fault models captures the fault.
\begin{figure}[tb]
	
	\centering
	\includegraphics[width=0.85\linewidth]{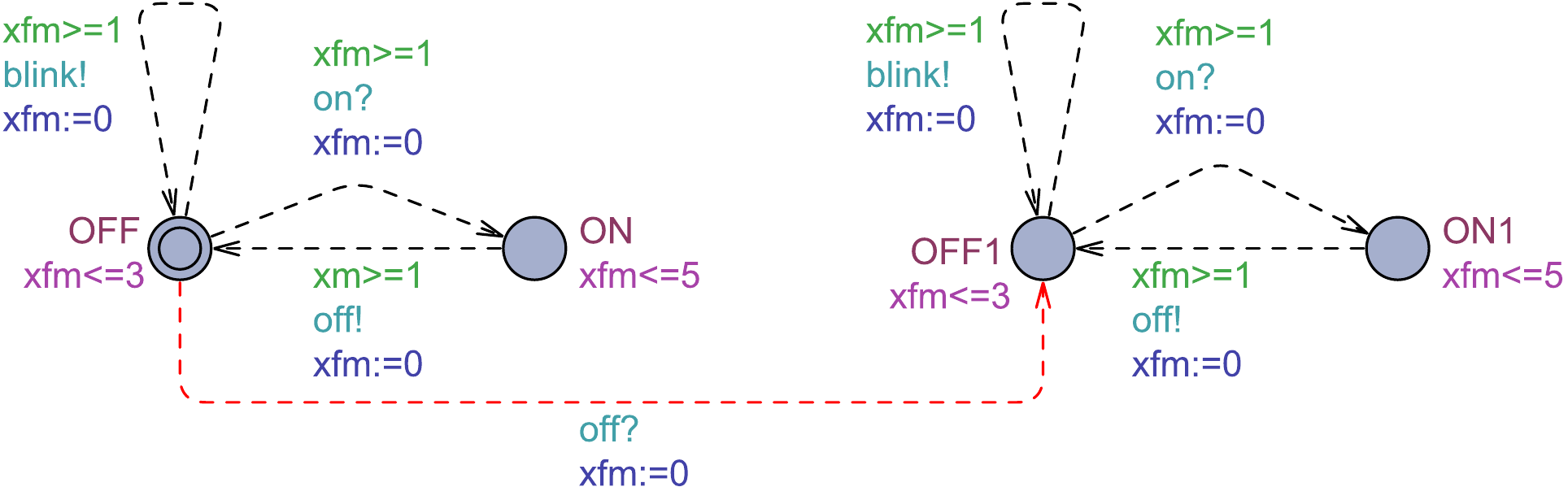}\\
	
	\caption{A fault model \fspeci for a transient fault that captures an unexpected output that resets a clock. }
	\label{fig:fault}
\end{figure}
\begin{definition}[Fault Model]
Let \spec be a specification. A fault model \fspeci for \spec consists of two copies of the specification \spec, one copy for the pre-fault behavior and one copy for the
post-fault behavior. The two copies are connected with a single
transient error.
\end{definition}
\textbf{Example.} \emph{Fig.~\ref{fig:fault} gives an example of a fault model. 
The fault captures the situation, in which an \uppSync{off} signal is produced in the \uppLoc{OFF} location and this faulty signal additionally resets the clock \uppClock{xfm}, but does not change the location of the automaton. In the fault model,
this transient fault leads from the \uppLoc{OFF} location from the pre-fault 
part to the \uppLoc{OFF} in the post-fault part.}

Since we cannot observe the internal state of the system, 
we do not know which fault model captured the fault that occurred. 
Thus, the shield can only end the recovery phase if all fault models and the specification \textit{align}; i.e.,
if all \fspeci and \mspec reach the same state.
To achieve this, in the recovery phase, we monitor the behavior of the system and update the fault models accordingly.
If a fault model can not follow the output of the system (including not
allowing a delay that is possible in the system), it was not the
correct fault model for the observed fault and is discarded.
Only if all \emph{active}, i.e., non-discarded, fault models agree on the same state, 
the shield and the system synchronized again.

\textbf{Types of transient faults.}
We consider fault models covering the following fault types; categorized in location-faults, clock-faults, and their combination.
\begin{itemize}
\item{Location-faults:}
\begin{itemize}
\item{\emph{Go-to-any-location} faults:} the system goes to an arbitrary location.
To track the its location, we need a fault model for every location in \spec.
\item{\emph{Go-to-next-location} faults:} the system gives an incorrect output,
but continues in a correct successor location. Thus, the fault models only need to cover the valid successor locations.
\end{itemize}
\item{Clock-faults:}
\begin{itemize}
\item{\emph{Wrong-reset} faults:}  the system illegally resets a clock to zero.
Such faults occur, e.g., if  the system gives a planned reset at a wrong point in time, or the systems resets the wrong clock.
\item{\emph{Swapped-clocks} faults:} the system swaps the values of several selected clocks. This might be a {\em binary} swaps between two clocks, or {\em permutations} between several clocks.
\item{\emph{Missing-reset} faults:} the system ignores a planned reset of a clock, resulting in a clock value that is too high.
\end{itemize}
\end{itemize}

\textbf{Example.} \emph{The fault shown in Fig.~\ref{fig:fault}(a) is a wrong-reset fault, i.e., the fault does not change the location of the model, but only resets the clock \uppClock{xfm}.}

\subsection{Definition of Timed Post-Shields with Recovery Guarantees}

We now define shields which satisfy an additional property:  
\textit{guaranteed-recovery} or \textit{guaranteed-time-bounded-recovery}.

\textbf{The Guaranteed-Recovery Property.} 
In this paper, we synthesize shields with guaranteed (time-bounded) recovery 
under the assumption that the system satisfies the specification except for a single transient fault and that this fault is covered by one of the fault models \fspeci with $i\in \{1\dots n\}$.

\begin{definition}[Guaranteed Recovery]
Let \spec be a specification, let \mspec be its monitor, let $\text{\fspec}=\{\text{\fspecone}\dots\text{\fspecn}\}$ be a set of fault models, let \sys be a timed system with $\sys \leq$ \fspeci for some \fspeci $\in$ \fspec. We say that \sh guarantees recovery if for every trace $\sigma$ containing a single transient fault, i.e.,
there exists a point in time $t_1$ such that  \mspec reaches the error location \uppLoc{ERR} at $t_1$,
there exists a time $t_2>t_1$ such that after $t_2$ $(\sys | \sh)$ keeps $\sigma$ intact. 
\end{definition}
That is, if the system refines any of our considered fault models, we guarantee that an observed error will lead to recovery and the system and the shield give the same output again.  

\textbf{The Guaranteed-Time-Bounded-Recovery Property.} 
This property guarantees that the recovery phase lasts for at most $T$ time units
after a fault.

\begin{definition}[Guaranteed Time-Bounded Recovery]
Let \spec be a specification, let \mspec be its monitor, let $\text{\fspec}=\{\text{\fspecone}\dots\text{\fspecn}\}$ be a set of fault models, let \sys be a timed system with $\sys \leq$ \fspeci for some \fspeci $\in$ \fspec. We say that \sh guarantees recovery within a bound $T$ if for every trace $\sigma$ containing a single transient fault, i.e.,
there exists a point in time  $t_1$ such that  \mspec reaches the error location \uppLoc{ERR} at $t_1$,
we have that after time $t_1+T$, $(\sys | \sh)$ keeps $\sigma$ intact. 
\end{definition}

\subsection{Construction of Timed Post-Shields with Recovery Guarantees}

In this section we discuss the synthesis procedure of timed timed post-shields with 
\textit{guaranteed recovery} and with \textit{guaranteed time-bounded recovery}.

\textbf{Algorithm 2.} Let \spec be a specification, \mspec its monitor, 
and 
$\text{\fspec}=\{\text{\fspecone}\dots\text{\fspecn}\}$
a set of fault models.
Starting from \spec, \mspec, \fspec , we construct a timed post-shield 
with guaranteed-recovery (\shg), or 
with guaranteed-time-bounded-recovery (\shgt) via the following steps. 

\textbf{Steps 1 and 2.} Perform as in Section 3.2.

\textbf{Step 3: Construct the monitors \mfspeci}.
Transform the fault models \fspeci into monitors \mfspeci for $i\in \{1\dots n\}$.

\textbf{Step 4: Construct the timed Game $\mathcal{G}$.}
Now we can consider the timed game given by the following composition:
$$\mathcal{G} = \text{\mfspecone}~|...|~\text{\mfspeci}~|~\mspec~|~\mspecp~ |~\shield$$
In this game, we observe conformance of the system with respect
to the fault models \mfspeci and the original specification via \mspec,
enforce the correctness of the shield via \mspecp, and ensure
no unnecessary deviations via \shield.
Next, we compute the winning strategies $\omega_g$ and $\omega_{gt}$ of $\mathcal{G}$ such that that the corresponding shields guarantee recovery and guarantee time-bounded recovery, respectively.

\textbf{Step 5a: Compute a strategy $\omega_g$ of $\mathcal{G}$ for guaranteed-recovery.}
For \emph{guaranteed recovery}, we need to establish a state where all
active 
fault models agree with each other and the \specp, that is, they are all in the same location with same clock values. This can be achieved with the following leads-to property~\cite{behrmann2004tutorial}, specifying that if we observe a fault, this will eventually lead to recovery. 
\begin{align*}
	&\uppProp{control: A \Box~ \neg \mspecp.\uppLoc{ERR} \wedge \mspec.\uppLoc{ERR}~leadsto} \\ 
	&\uppProp{(\forall i. [\mspec^{f_i}.\uppLoc{ERR}~ \vee~ (\mspec^{f_i}.\uppVar{l} == \mspecp.\uppVar{l} ~\wedge
	~\mspec^{f_i}.\uppClock{x} = \mspecp.\uppClock{x})])}
\end{align*}

Solving $\mathcal{G}$ w.r.t. this query will produce a strategy $\omega_g$.
Note, that if a fault model is inactive, i.e., it reached its error-state, this means that \sys performed an output that was not valid in the fault model. 

\textbf{Step 5b: Compute a strategy $\omega_{gt}$ of $\mathcal{G}$ for guaranteed-time-bounded-recovery.}
We slightly change the  timed leads-to property to compute  $\omega_{gt}$
for guaranteed recovery within a time bound $T$:
\begin{align*}
	&\uppProp{control: A\Box~\neg \mspecp.\uppLoc{ERR} \wedge \mspec.\uppLoc{ERR} ~leadsto_{\leq T}} \\ 
	&(\forall i . [\mspec^{f_i}.\uppLoc{ERR} ~\vee~ (\mspec^{f_i}.\uppVar{l} == \mspecp.\uppVar{l} ~\wedge
	~\mspec^{f_i}.\uppClock{x} = \mspecp.\uppClock{x})])
\end{align*}
\textbf{Step 6: Construction of the timed post-shields \shg and \shgt.}
We construct a shield with guaranteed-recovery \shg by $\mathcal{G}\|\omega_g$,
and a shield with guaranteed-time-bounded-recovery \shgt by $\mathcal{G}\|\omega_{gt}$.

\begin{theorem}
The shields \shg and \shgt, constructed by Alg. 2, are timed post-shields with guaranteed-recovery and guaranteed-time-bounded-recovery, resp.
\end{theorem}
\emph{Proof.} Correctness and no-unnecessary-deviation are given as for regular timed post-shields. \shg (\shgt) is a shield with guaranteed-(time-bounded)-recovery, simply by the query it was produced from, which can only be satisfied if all traces in  \shg (\shgt) either do not encounter a fault, or the fault will lead to recovery at some point (within T time units) along the trace.  \qed

\textbf{Discussion: timed post-shields.}
Both types of shields, with and without recovery, have pros and cons.
Obviously, the ability of guaranteed recovery after a system fault is highly desirable.
This is especially true, if the system to be shielded is highly optimized and
performs complex tasks that are not captured in the specification of the shield.
Nevertheless, it may not be feasible to synthesize post-shields with the ability to recover.
First, guaranteed recovery is not always possible and therefore, no shield that guarantees recovery may exist. An obvious example that demonstrates this fact is a shield for a system that never resets one of its clocks. If a fault occurred that changes the value of this clock, synchronization is never possible. 
Second, to capture any transient fault that is possible under a given fault category, an exponential number of fault models is needed. This results in an exponential blowup in the state space and synthesis time. Therefore, it may not be feasible to synthesize
shields with guaranteed recovery for large specifications considering a large number of fault models. 

%% file: preempt.tex
In this section we define and construct timed pre-shields.
A timed pre-shield is attached before the system as illustrated in Figure~\ref{fig:attach_shield}(a).
At any point in time, a timed pre-shield provides \emph{a set of actions} \act
for the system to choose from.
This set of action can contain a delay action.
If this is the case, the system is permitted to wait without performing any discrete action. If the set does not include a delay action, the system has to produce an output immediately.
If the system picks the outputs according 
to this list, then it is guaranteed that the shield and the 
timed system together satisfy the specification.

\subsection{Definition of Timed Pre-Shields}

In this section we define timed pre-shields based on two properties;
correctness and no unnecessary restriction.

\begin{definition}[Correctness for Pre-Shields]
\label{def:corr_pre}
Let \spec be a specification, and let \sh be a pre-shield.
For any state $s$ from \sh, let \act$={\alpha_1,\dots,\alpha_n}$ be the set of enabled 
actions sent by \sh.
A pre-shield is correct, if it holds that if $\alpha$
is a wrong action for the given situation, then $\alpha\not\in\act$.
\end{definition}

\begin{definition}[No-Unnecessary-Restriction]
\label{def:mo_per}
Let \spec be a specification, and let \sh be a pre-shield.
For any state $s$ from \sh, let \act$={\alpha_1,\dots,\alpha_n}$ be the set of enabled 
actions sent by \sh.
A pre-shield is not unnecessarily restrictive, if it holds that if $\alpha$
is a correct action for the given situation, then $\alpha\in\act$.
\end{definition}

\begin{definition}[Timed Pre-Shields.]
Given a specification \spec. \sh is a timed pre-shield if it holds that \sh enforces correctness for any timed system \sys w.r.t. \spec (Def.~\ref{def:corr_pre}) and \sh is not unnecessarily restrictive (Def.~\ref{def:mo_per}).
\end{definition}
 
\subsection{Construction of Timed Pre-Shields}
We construct timed pre-shields according to the following algorithm.

\textbf{Algorithm 3.}
Starting from \spec, \mspec, we construct a timed pre-shield. 

\textbf{Step 1. Construction of the automaton \shield.} 
Since there is no concept of minimal deviation in pre-shields, the control options of the \shield component are not restricted. Instead, in this setting, we build \shield such that it can fire any output at any time, i.e., it has a single location and unguarded self-loops for every output. Note that timed pre-shields do not need primed outputs. 

\begin{figure}[tb]
	
	\centering
	\includegraphics[width=0.35\linewidth]{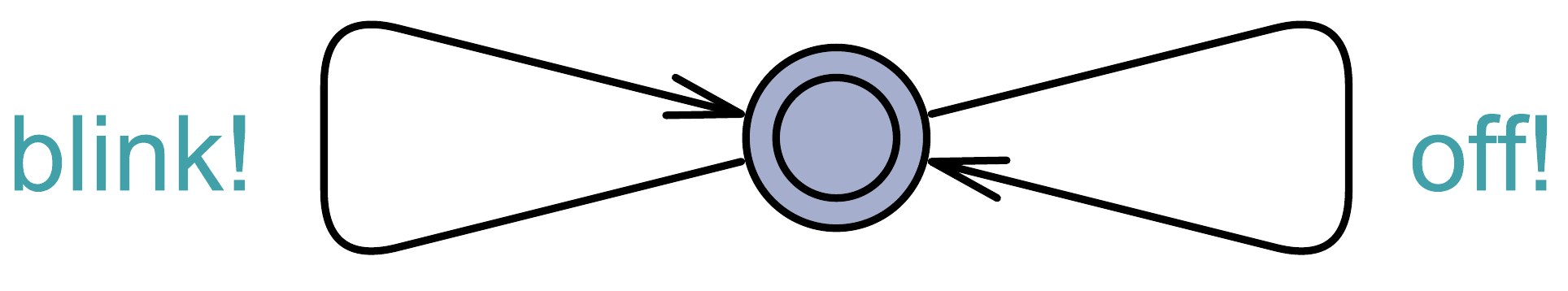}\\
	
	\caption{ The control options for a timed-pre shield of the light switch. }
	\label{fig:ctr_n}
	\vspace{-10pt}
\end{figure}

 \textbf{Example.} \emph{The component \shield for the light switch is depicted in Fig.~\ref{fig:ctr_n}}. 

\textbf{Steps 2. Construct the timed safety game $\mathcal{G}$.}  
We construct the timed game $\mathcal{G}$ by the following composition:
$$\mathcal{G} = ~\mspec~|~\shield$$

\textbf{Step 3: Compute a strategy $\omega_{pre}$ of $\mathcal{G}$.}
Pre-shields ensure correctness (safety), i.e., 
the control objective is to ensure that \spec is
never violated. Solving the safety game w.r.t. the following query produces a strategy $\omega_{pre}$.
$$
\uppProp{control: A \Box~ \neg \mspec.\uppLoc{ERR}}
$$

\textbf{Step 4: Computing the set of enabled actions via zones.}
For a given state, the set of enabled actions \act is computed via zones.
From any given state, its zones can be calculated straightforwardly, see~\cite{alur1994theory}.
\begin{figure}[tb]
\centering
\resizebox{0.38\textwidth}{!}{

\tikzset{every picture/.style={line width=0.75pt}} 

\begin{tikzpicture}[x=0.75pt,y=0.75pt,yscale=-1,xscale=1]

\draw  [draw opacity=0] (100,32.33) -- (370.83,32.33) -- (370.83,272) -- (100,272) -- cycle ; \draw  [color={rgb, 255:red, 180; green, 180; blue, 180 }  ,draw opacity=1 ] (100,32.33) -- (100,272)(130,32.33) -- (130,272)(160,32.33) -- (160,272)(190,32.33) -- (190,272)(220,32.33) -- (220,272)(250,32.33) -- (250,272)(280,32.33) -- (280,272)(310,32.33) -- (310,272)(340,32.33) -- (340,272)(370,32.33) -- (370,272) ; \draw  [color={rgb, 255:red, 180; green, 180; blue, 180 }  ,draw opacity=1 ] (100,32.33) -- (370.83,32.33)(100,62.33) -- (370.83,62.33)(100,92.33) -- (370.83,92.33)(100,122.33) -- (370.83,122.33)(100,152.33) -- (370.83,152.33)(100,182.33) -- (370.83,182.33)(100,212.33) -- (370.83,212.33)(100,242.33) -- (370.83,242.33) ; \draw  [color={rgb, 255:red, 180; green, 180; blue, 180 }  ,draw opacity=1 ]  ;
\draw   (130,121.33) -- (250,121.33) -- (250,211.33) -- (130,211.33) -- cycle ;
\draw   (220,91.33) -- (340,91.33) -- (340,181.33) -- (220,181.33) -- cycle ;
\draw    (160,241.33) -- (370,32.33) ;
\draw [shift={(160,241.33)}, rotate = 315.14] [color={rgb, 255:red, 0; green, 0; blue, 0 }  ][fill={rgb, 255:red, 0; green, 0; blue, 0 }  ][line width=0.75]      (0, 0) circle [x radius= 3.35, y radius= 3.35]   ;
\draw   (161.83,240.33) .. controls (165.08,243.69) and (168.38,243.75) .. (171.73,240.51) -- (174,238.32) .. controls (178.8,233.69) and (182.82,233.05) .. (186.06,236.41) .. controls (182.82,233.05) and (183.6,229.06) .. (188.39,224.43)(186.23,226.51) -- (190.66,222.23) .. controls (194.02,218.99) and (194.08,215.69) .. (190.84,212.34) ;
\draw   (219.83,181.33) .. controls (216.59,177.98) and (213.29,177.92) .. (209.94,181.17) -- (207.16,183.85) .. controls (202.37,188.48) and (198.35,189.12) .. (195.1,185.77) .. controls (198.35,189.12) and (197.57,193.12) .. (192.78,197.75)(194.94,195.67) -- (190,200.44) .. controls (186.65,203.68) and (186.59,206.98) .. (189.83,210.33) ;
\draw   (219.83,181.33) .. controls (223.13,184.63) and (226.43,184.63) .. (229.73,181.33) -- (232.71,178.35) .. controls (237.42,173.64) and (241.43,172.93) .. (244.73,176.23) .. controls (241.43,172.93) and (242.14,168.92) .. (246.85,164.21)(244.73,166.33) -- (249.83,161.23) .. controls (253.13,157.93) and (253.13,154.63) .. (249.83,151.33) ;
\draw   (309.83,91.33) .. controls (306.53,88.03) and (303.23,88.03) .. (299.93,91.33) -- (281.95,109.31) .. controls (277.24,114.02) and (273.23,114.73) .. (269.93,111.43) .. controls (273.23,114.73) and (272.52,118.74) .. (267.81,123.45)(269.93,121.33) -- (249.83,141.43) .. controls (246.53,144.73) and (246.53,148.03) .. (249.83,151.33) ;
\draw   (310,91) .. controls (313.21,94.39) and (316.51,94.47) .. (319.9,91.26) -- (338.48,73.63) .. controls (343.32,69.04) and (347.34,68.44) .. (350.55,71.83) .. controls (347.34,68.44) and (348.16,64.46) .. (352.99,59.87)(350.82,61.93) -- (371.58,42.23) .. controls (374.97,39.02) and (375.05,35.73) .. (371.84,32.34) ;
\draw    (101,272) -- (370.83,272.33) ;
\draw [shift={(372.83,272.33)}, rotate = 180.07] [color={rgb, 255:red, 0; green, 0; blue, 0 }  ][line width=0.75]    (10.93,-3.29) .. controls (6.95,-1.4) and (3.31,-0.3) .. (0,0) .. controls (3.31,0.3) and (6.95,1.4) .. (10.93,3.29)   ;
\draw    (101,272) -- (100.01,34.33) ;
\draw [shift={(100,32.33)}, rotate = 449.76] [color={rgb, 255:red, 0; green, 0; blue, 0 }  ][line width=0.75]    (10.93,-3.29) .. controls (6.95,-1.4) and (3.31,-0.3) .. (0,0) .. controls (3.31,0.3) and (6.95,1.4) .. (10.93,3.29)   ;

\draw (82,33) node [anchor=north west][inner sep=0.75pt]   [align=left] {\LARGE y};
\draw (375,252) node [anchor=north west][inner sep=0.75pt]   [align=left] {\LARGE x};
\draw (133,102) node [anchor=north west][inner sep=0.75pt]   [align=left] {\LARGE a};
\draw (220,73) node [anchor=north west][inner sep=0.75pt]   [align=left] {\LARGE b};
\draw (187,234) node [anchor=north west][inner sep=0.75pt]   [align=left] {\LARGE z1};
\draw (178,170) node [anchor=north west][inner sep=0.75pt]   [align=left] {\LARGE z2};
\draw (253,165) node [anchor=north west][inner sep=0.75pt]   [align=left] {\LARGE z3};
\draw (252,95.33) node [anchor=north west][inner sep=0.75pt]   [align=left] {\LARGE z4};
\draw (353,68) node [anchor=north west][inner sep=0.75pt]   [align=left] {\LARGE z5};
\draw (140,233) node [anchor=north west][inner sep=0.75pt]   [align=left] {\LARGE S};

\end{tikzpicture}
}
\caption{The zones reachable by delay from a state $s$ in a TIOA with two clocks, \uppClock{x} and \uppClock{y}. The squares represent  constraints for two transitions with label \uppSync{a} and \uppSync{b}.}
\label{fig:regions}
\end{figure}
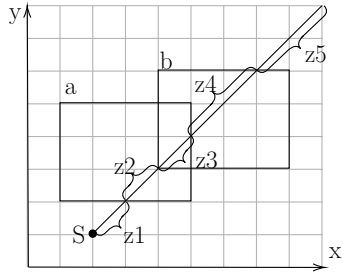

 \textbf{Example.} \emph{The concept of zones is illustrated in Fig.~\ref{fig:regions}, where the $X$ and $Y$ axis depict different clocks, and the squares represent the constraints in which different controllable actions $a$ and $b$ are enabled. We have for $a: \{1<x<5, 2<y<5\}$ and for $b: \{4<x<8, 3<y<6\}$}. 

The set of enabled actions is kept up to date by monitoring the current state of $\sh_{pre}$.
 Whenever a new input or output from the system is received, the state is updated. From the new state, we calculate all zones that can be reached via delay and the actions enabled in each zone.  The set of actions \act for the current zone is sent to the system. In case the end of the zone is not met yet, this includes a delay action. If enough time passes so that the end of the current zone is met, the shield needs to check whether future zones permit actions. If so, the set of actions of the next zone is passed to the system. Otherwise, the current set of actions is transmitted again, this time without a delay action. The system may choose any of the permitted actions, including delay if possible.

 \textbf{Example.} \emph{In Fig.~\ref{fig:regions}, the actions enabled per zone are $z1=\{\}, z2=\{$\uppSync{a}$\}, z3=\{$\uppSync{a},\uppSync{b}$\}, z4=\{$\uppSync{b}$\}, z5=\{\}$. Thus, in the state $S$ which is in $z1$, \act is empty, after one time unit $\act=\{a\}$, and so on.} 



\begin{theorem}
A shield $\sh_{pre}$ constructed according to Alg. 3 is a timed pre-shield.
\end{theorem}

\emph{Proof}. In order for $\sh_{pre}$ to be a timed pre-shield, $\sh_{pre}$ needs to satisfy correctness for pre-shields and the no-unnecessary-restriction property. $\sh_{pre}$ is correct, as the safety game removes every transition leaving the winning region. It has no-unnecessary-restriction, as \shield is not restricted when producing outputs, and all correct actions are kept when producing the strategy. \qed

\tikzset{every picture/.style={line width=0.75pt}} 

\textbf{Discussion: Pre-shielding vs. Post-shielding.}
Post-Shielding has the advantage, that we treat the system as a total back box.
In order for pre-shielding to work, we need a system that chooses the actions
w.r.t. the suggestions of the shield. Usually, we have this setting in RL
where the shield can easily influence the set of available actions from the agent.
Instead of overruling the system, a pre-shield 
leaves the choice of which action should be executed always to the system, i.e.,
the system can do anything as long as it is safe. 
Thus, the overall efficiency of the system is kept as high as possible. 
It has already been shown that removing unsafe options during learning can significantly speed up the learning process~\cite{AlshiekhBEKNT18}.

	
	
	

%% file: experiments.tex
To validate our approach, we extend the case-study of Larsen et. al.~\cite{LarsenMT15} to a platoon of multiple cars\footnote{The source code, including some demonstrative videos and the running example used in the paper, is available online \cite{alexander_palmisano_2020_3903227}.}.
In the case study, an RL agent controls $n$ \emph{follower} vehicles in the platoon following an (environment controlled) \emph{leader} vehicle. 
All vehicles can drive a maximum of $20$m/s and have three different possible accelerations modes: $-2$m/s$^2$, $0$m/s$^2$ and $2$m/s$^2$ which can be changed at every time unit.
The goal of the RL agent is to control the followers in the platoon
such that the total distance between all vehicles is minimized.
Furthermore, the RL agent receives a negative reward if the distance between two cars is outside a safe region ($\leq 5$m) or is too large (above $200$m).
The hyper-parameters of the RL setting can be found in Appendix~\ref{app:RL}.
We used the models from~\cite{LarsenMT15} and synthesized timed post-shields with \uppaaltiga, as discussed in Sec.~\ref{sec:ts}.
We study the behaviour of RL agents in the context of 
\begin{enumerate*}
    \item no shielding,
    \item post-shielding during execution, and
    \item post-shielding during both training and execution.
\end{enumerate*}
We report the learning curves during the training phase and the performances in the execution phase for $n\in\{2,4,6,8,10\}$ where $n$ denotes the number of cars.
\begin{figure}[t]
	\begin{center}
		\includegraphics[width=1\linewidth]{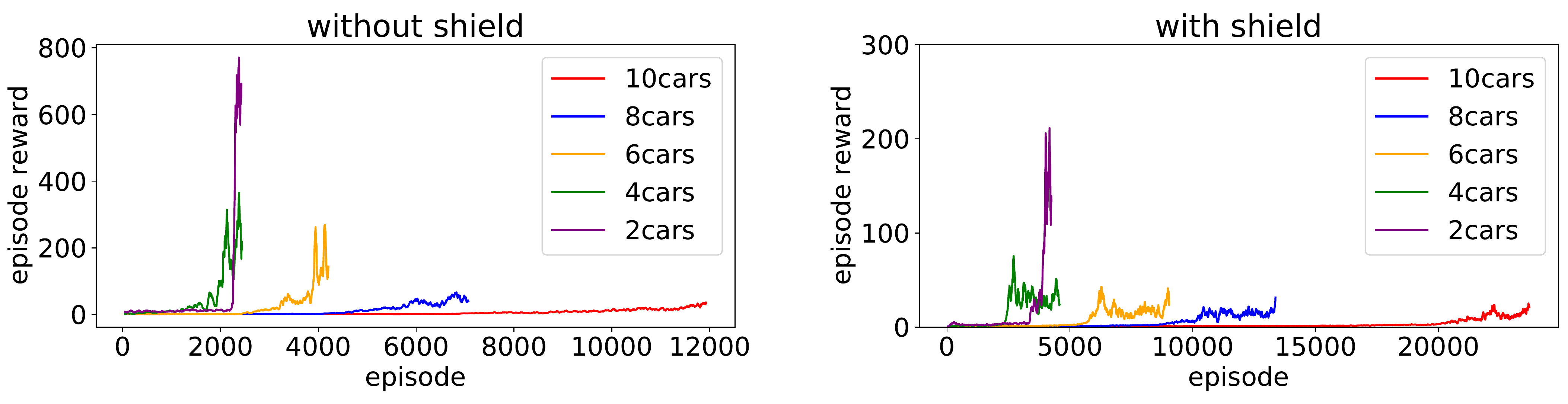}
		\caption{Results training phase.}
		\label{fig:result_training}
	\end{center}
	\vspace{-10pt}
\end{figure}

\emph{Training.}
Each training episode starts with random but safe initial distances and velocities of all cars.
During the simulation, the environment picks the accelerations of the leading car via a uniform distribution. 
A training episode lasts for 2000 time units, or until the distance between two cars gets smaller than $5$m  or larger than $200$m. 
Note, that with a shield, a training episode always lasts $2000$ time units, since safety is always guaranteed.

Fig.~\ref{fig:result_training} compares the learning curves as a mean of $20$ training phases, for the unshielded case (left) and the shielded case (right). 
The reward in the unshielded case is considerably higher than in the shielded setting.
We observe that the agent exploits the relatively low risk of a crash and makes potentially unsafe choices.
Since the accelerations of the leading car are picked via an uniform distribution, it is unlikely, that e.g., the leading car accelerates to the maximum speed and then immediately hits the break until it reaches zero. 
Such risk tolerance is not allowed when deploying the shield as even a potential but unlikely future crash should be shielded against.
\emph{Execution Phase.}
We tested all controller combinations for $1000$ simulations, and each simulation lasts until a crash or for a maximum of $2000$ time units.
 Table~\ref{tab:results_execution} depicts the results. Note, that we learned a global controller for each number of cars (but use local shields) and that the controllers optimize a local minimum, therefore the controllers performances differ from each other.
Interestingly, we observe that the combination of unshielded training (Shield E) provides better results in our setting, compared to a RL agent utilizing the shield also during training (Shield T+E).
But more experiments are needed to discern this effect in more detail.

\begin{table}[tb]
\begin{center}
 \begin{tabular}{|c | c c c |c| c|}
\hline
&\multicolumn{3}{|c|}{No Shield} &\;Shield E\;&\;Shield T+E\;\\\hline
\;\#Cars\;&\;\#Crashes\;&\;Time\;&\;Reward\;&\;Reward\;&\;Reward\; \\\hline
2 & 703 & 1133 & 747 & 915 & 603 \\ \hline
4 & 13 & 1989 & 1070 & 685 & 393 \\ \hline
6 & 0 & 2000 & 638 & 617 & 375 \\ \hline
8 & 85 & 1908 & 477 & 495 & 386 \\ \hline
10 & 983 & 544 & 170 & 608 & 342 \\ \hline

\end{tabular}
  \caption{Results exploitation phase using 10000 simulations. 
  Number of crashes is given in absolute values over all simulations whereas Reward and Time measures are given as averages.
  Time and Crash values are omitted when shielding is applied as these are $>2000$ and 0 respectively. Time denotes the time-units of simulation prior to a first crash.}
  \label{tab:results_execution}
\end{center}
	\vspace{-0.5cm}
\end{table} 

%% file: conclusion.tex
 We presented timed post-shields and timed pre-shields. 
 These are shields for real-time systems which can be attached either before or after a system, correcting the outputs received by the environment. 
 In addition, we discussed how timed shields can be used in reinforcement learning settings. 
 We presented a case study of a platoon of cars, and demonstrated the potential of a timed post-shield in RL. 
In the future, we would like to extend this work into several different directions.
First, we see great potential for shields
 in RL. In future work, we want to apply timed shields
on several challenging RL settings and investigate techniques for speeding up the learning performance in addition to providing safety.
Furthermore, we want to exploit techniques from model repair and model refinement
to deal with dynamic environments, and adapt the shields during runtime if needed.
In this work, we treat the system as adversary. In future work, we plan to study ways to model the spectrum between cooperative and adversarial systems to be shielded.

